\documentclass[twocolumn,a4paper,11pt]{article}
\usepackage{geometry}
\usepackage{amsmath,amssymb}
\usepackage{bm,braket}
\usepackage{multicol}
\usepackage{comment}
\usepackage[dvipdfmx]{graphicx}
\usepackage{cite}
\usepackage{authblk}

\title{{\Large Quantum phase transition between spin liquid and spin nematics in spin-1 Kitaev honeycomb model}}
\author{Tohru Mashiko and Tsuyoshi Okubo}
\affil{Institute for Physics of Intelligence, University of Tokyo, Tokyo 113-0033, Japan}
\date{}

\begin{document}
\small
\twocolumn[
\maketitle
\begin{quotation}
Besides the exactly solvable spin-1/2 Kitaev model, higher spin-$S$ ones, not exactly solvable, are promising playgrounds for researches on the quantum spin liquid as well.
One of the main interests in higher spin-S cases is the interplay between the Kitaev spin liquid (KSL) and spin nematics.
We probe this interplay in a spin-1 model on the honeycomb lattice with competing bilinear-biquadratic and Kitaev interactions.
Utilizing the 2D infinite projected entangled-pair state (iPEPS), we map out the phase diagram for the ferro-biquadratic interaction.
In the phase diagram, we discover the direct KSL--spin-nematics transitions in the vicinity of pure Kitaev limits.
It has been revealed that the ferro KSL exhibits robustness against perturbations from ferro-quadrupolar interactions.
Also, the spin-nematic phase is extended to the parameter region near the antiferro-Kitaev limit. \\
\end{quotation}
]

\section{Introduction}
\label{sec:intro}
The exactly solvable spin-1/2 Kitaev honeycomb model (KHM) exhibits \cite{kitaev} quantum spin liquids as the ground state, called the Kitaev spin liquid (KSL), ascribed for the bond-dependent spin-spin interactions giving rise to the frustration.
This KHM possesses the $\mathbb{Z}_{2}$ gauge structure of the Majorana fermions, i.e., the conservation of the flux, on each local hexagonal plaquette.
The Kitaev interactions can be realized in materials with strong spin-orbit couplings (SOCs) \cite{jackeli}.
There appeared a number of successive experiments with candidate Kitaev materials [e.g., $\alpha$-RuCl$_{3}$ \cite{kasahara,plumb,banerjee,winter,little,wang,ran}, $A_{2}$IrO$_{3}$ ($A=$Na, Li) \cite{chaloupka1,chaloupka2,chun,choi,singh1,singh2,katukuri,williams,rau} and H$_{3}$LiIr$_{2}$O$_{6}$ \cite{bette}], which then motivated researchers for more theoretical investigations \cite{kimchi,nasu2,yoshitake,catuneanu,gohlke,lee,leekaneko,zhang2}.

Besides the spin-1/2 KHM, higher spin-$S$ ones have also attracted attentions, despite the difficulties due to the absence of exact solutions.
Analytical studies of arbitrary spin-S Kitaev models have confirmed \cite{baskaran2} the conservation of $\mathbb{Z}_{2}$ gauge flux, similar to the case of spin-1/2, and the disappearance of spin-spin correlations beyond nearest neighbors.
Several numerical studies supported the existence of quantum spin liquids in spin-1 KHMs \cite{oitmaa,koga,lee2,dong,khait,zhu}.
Candidate materials were also proposed in spin-1 \cite{stavropoulos} and spin-3/2 \cite{leea,xu} cases.

Higher internal degrees of freedom than spin-1/2 cause unique spin-quadrupolar (or multipolar) states.
One of the most representative ones is the spin nematic (SN) state characterized by the spin-quadrupolar order.
The interplay between the quantum spin liquid and spin nematic state has been regarded as a significant topic, which can not be observed in the classical system but in the quantum system.
One may discover novel quantum properties of matters by investigating this interplay on the KHM whose ground state is verified to be the quantum spin liquid.
However, this is far from understood since it is hard to investigate their ``hidden'' magnetism.
Namely, both states do not exhibit the long-range spin-dipolar order, although they are different in that the SN breaks spin rotational symmetry owing to the order of the spin-quadrupolar moments \cite{blume,matveev,tsunetsugu,lauchli1,penc} while quantum spin liquids including the KSL do not spontaneously break any symmetries \cite{lacroix}.
\begin{figure*}[t]
    \begin{center}
    \includegraphics[keepaspectratio,scale=0.52]{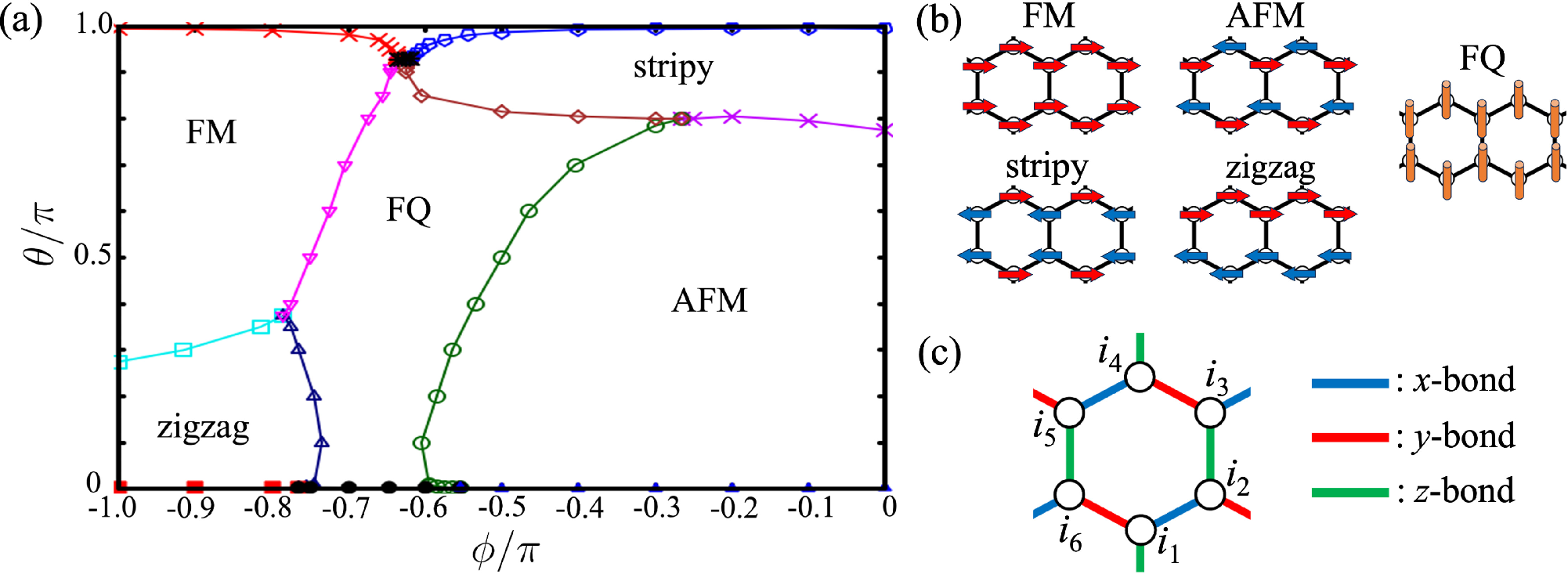}
    \end{center}
        \caption{(a) Phase diagram of the BBQ-K model in the region of $0.0\le\theta/\pi\le1.0$ and $-1.0\le\phi/\pi\le0.0$. 
        (b) Configurations of FM, AFM, stripy, zigzag, and FQ phases. 
        The arrows and rods denote spin dipolar- and quadrupolar-moments respectively. 
        (c) Hexagonal plaquette with sites $\{i_{1},i_{2},\cdots,i_{6}\}$. 
        The blue, red and green lines denote bonds of local tensors along $x$-, $y$-, and $z$-directions of the Kitaev term respectively.}
        \label{fig:phase_etc}
\end{figure*}

Recently, Pohle et al. addressed \cite{pohle} this challenging problem by investigating a spin-1 honeycomb model with competing bilinear-biquadratic (BBQ) and Kitaev interactions, which we call the BBQ-K model defined as
\begin{eqnarray}
    \hat{H}_{\mathrm{BBQ-K}} &\equiv& \sum_{\gamma=x,y,z} \sum_{\langle i,j \rangle_{\gamma}} \hat{H}^{\langle i,j \rangle_{\gamma}}_{\mathrm{BBQ-K}}, \label{bbqk} \\ 
    \hat{H}^{\langle i,j \rangle_{\gamma}}_{\mathrm{BBQ-K}} &\equiv& J_{1} \hat{\bm{S}}_{i} \cdot \hat{\bm{S}}_{j} + J_{2} \left( \hat{\bm{S}}_{i} \cdot \hat{\bm{S}}_{j} \right)^{2} + K \hat{S}^{\gamma}_{i} \hat{S}^{\gamma}_{j}. \label{bbqkij}
\end{eqnarray}
Here, $\langle i,j \rangle_{\gamma}$ denotes the nearest neighbor pair on a $\gamma$ bond.
$J_{1}$, $J_{2}$, and $K$ are coupling coefficients of Heisenberg (bilinear), biquadratic, and Kitaev terms respectively.
They discovered a variety of quantum properties caused by the competition between these interactions.
However, several properties including phases and phase transitions are still unclear when the Kitaev interaction prevail over the BBQ ones.
This is because it is in principle impossible for semi-classical variational calculations adopted in Ref. \cite{pohle} to illustrate quantum spin liquid.

In this paper, we investigate the BBQ-K model to explore quantum phases and phase transitions, extended to the parameter regions where the Kitaev interaction is dominant.
We here adopted the tensor network method, that is the 2D infinite projected entangled-pair state (iPEPS) \cite{verstraete2,verstraete3,jordan}, to accurately capture the quantum entanglements among spins, which was neglected in Ref. \cite{pohle}.

As a result, we obtain the phase diagram Fig. \ref{fig:phase_etc}(a).
Following the notation in Ref. \cite{pohle}, we normalize the coefficients in Eq. \eqref{bbqkij} with two parameters $\theta$ and $\phi$ as $( J_{1}, J_{2}, K ) \equiv ( \sin\theta \cos\phi, \sin\theta \sin\phi, \cos\theta )$.
Similar to Ref. \cite{pohle}, Fig. \ref{fig:phase_etc}(a) depicts four spin-dipolar ordered phases: antiferromagnetic (AFM), ferromagnetic (FM), zigzag, stripy phases, and the ferro-quadrupolar ordered spin nematic (FQ) phase [see Fig. \ref{fig:phase_etc}(b)].
However, unlike Ref. \cite{pohle}, strong competition between the Kitaev and ferro-biquadratic (ferro-quadrupolar) terms causes unconventional quantum properties: (1) two extended KSL phases and (2) the direct KSL--FQ transitions.
We describe details of these results in Sec. \ref{sec:results}.

This paper is organized as follows.
In Sec. \ref{sec:bbqk}, we describe the BBQ-K model.
Then, we show the details of the tensor network method in Sec. \ref{sec:tn}.
We present our numerical results: detailed phase diagrams and how to detect phases and boundaries, in Sec. \ref{sec:results}.
Then we discuss the nature of BBQ-K model in Sec. \ref{sec:con}, including implications for experiments to capture the phase diagram.

\section{BBQ-K Model}
\label{sec:bbqk}
In this section, we describe the BBQ-K model Eqs. \eqref{bbqk} and \eqref{bbqkij} in more detail.
At the limit of $\theta=0$ ($\pi$) [$J_{1}=J_{2}=0$ and $K=1$ ($-1$)], Eq. \eqref{bbqk} is nothing but the spin-1 Kitaev model whose ground state is antiferro- (ferro-) KSL, which we call the AKSL (FKSL).
These Hamiltonians commute \cite{baskaran2} with the $\mathbb{Z}_{2}$ gauge flux operator $\hat{W}_{p}$ on a hexagonal plaquette with sites $\{i_{1},i_{2},\cdots,i_{6}\}$ shown in Fig. \ref{fig:phase_etc} (c), defined as
\begin{eqnarray}
    \hat{W}_{p} \equiv \hat{U}^{z}_{i_{1}} \hat{U}^{y}_{i_{2}} \hat{U}^{x}_{i_{3}} \hat{U}^{z}_{i_{4}} \hat{U}^{y}_{i_{5}} \hat{U}^{x}_{i_{6}}, \label{flux}
\end{eqnarray}
where $\hat{U}^{\gamma}_{i} \equiv \exp( i\pi \hat{S}^{\gamma}_{i})$ and $\gamma=x,y,z$.
These spin-1 KSL states are characterized by the vortex freeness, i.e., $\langle\hat{W}_{p}\rangle=+1$, according to a study of the spin-wave theory \cite{baskaran2} and numerical studies \cite{koga,lee2}.

The limit of $\theta=\pi/2$ ($K=0$) is the BBQ model.
The biquadratic term $( \hat{\bm{S}}_{i} \cdot \hat{\bm{S}}_{j} )^{2}$ can be decomposed with the spin-quadrupolar operator $\hat{\bm{Q}}_{i}$ at site $i$, as
\begin{eqnarray}
    \left( \hat{\bm{S}}_{i} \cdot \hat{\bm{S}}_{j} \right)^{2} = \frac{1}{2} \left( \hat{\bm{Q}}_{i}\cdot\hat{\bm{Q}}_{j} - \hat{\bm{S}}_{i} \cdot \hat{\bm{S}}_{j} \right) + \frac{4}{3}, \label{stoq}
\end{eqnarray}
where the term $\hat{\bm{Q}}_{i}\cdot\hat{\bm{Q}}_{j}$ stabilizes the spin nematic ground state.
$\hat{\bm{Q}}_{i}$ has five components: $(\hat{S}^{x}_{i})^{2} - ( \hat{S}^{y}_{i} )^{2}$, $\sqrt{3} [ (\hat{S}^{z}_{i} )^{2} - (2/3) ]$, $\hat{S}^{x}_{i} \hat{S}^{y}_{i} + \hat{S}^{y}_{i} \hat{S}^{x}_{i}$, $\hat{S}^{y}_{i} \hat{S}^{z}_{i} + \hat{S}^{z}_{i} \hat{S}^{y}_{i}$, and $\hat{S}^{z}_{i} \hat{S}^{x}_{i} + \hat{S}^{x}_{i} \hat{S}^{z}_{i}$.
If $J_{2}<0$, the FQ phase is stabilized \cite{zhao}.
If $J_{2}>0$, the antiferro-quadrupolar (AFQ) ordered phase appears in the classical case \cite{chen,papanicolaou}, whereas this quadrupolar order is melted \cite{zhao,corboz} by quantum fluctuations.

\section{Tensor network method}
\label{sec:tn}
In our numerical calculations, we implemented variational calculations by the iPEPS \cite{verstraete2,verstraete3,jordan}.
Here, we provide an overview of the algorithms.

In the case of the finite-size system, the ground state $\left| \Psi \right\rangle$ of a spin-1 system with $N$ sites can be expressed by direct products of local spin-1 states as
\begin{eqnarray}
    \left| \Psi \right\rangle \equiv \sum_{\{s_{i}=0,\pm1\}} \Psi^{s_{0}\cdots s_{N-1}} \left| s_{0} \right\rangle \otimes \cdots \otimes \left| s_{N-1} \right\rangle, \label{peps1}
\end{eqnarray}
where $s_{i}=0,\pm1$ denotes the quantum number of $\hat{S}^{z}_{i}$.
We approximate the ground-state wave function $\Psi^{s_{0}\cdots s_{N-1}}$ as a project entangled-pair state (PEPS) \cite{nishino,verstraete} with $N$ local tensors $\psi^{s_{i}}_{l_{i}m_{i}n_{i}}$ as
\begin{eqnarray}
    \Psi^{s_{0}\cdots s_{N-1}} \approx \operatorname{Tr} \left( \prod_{i=0}^{N-1} \psi^{s_{i}}_{l_{i}m_{i}n_{i}} \right), \label{peps2}
\end{eqnarray}
where $\operatorname{Tr}$ stands for the contraction of all virtual indices $\{l_{i}m_{i}n_{i}\}$ which correspond to edges of the honeycomb lattice.
We define the bond dimension of virtual indices as $D$.  
We consider the iPEPS by enforcing translational symmetry in the PEPS to calculate physical quantities in the thermodynamic limit.
We assume N-sublattice structure in a unit cell.
The iPEPS can express a larger space of quantum states for larger $D$ (arbitrary quantum states in the case of $D\to\infty$).
Since we here consider only short range interactions, we can assume that the entanglement entropy obeys the area law \cite{srednicki,xiang,calabrese}, which allows us to efficiently approximate the ground state even with the finite bond dimension $D$ \cite{orus}.

Next, we search for the objective ground state $\left| \Psi \right\rangle$ making use of the imaginary-time evolution (ITE) as  $\left| \Psi \right\rangle \simeq e^{-T\hat{H}} \left| \Psi_{0} \right\rangle$, where $T$ is a sufficiently long imaginary time and $\left| \Psi_{0} \right\rangle$ is an initial state.
By the 1st order Trotter-Suzuki decomposition \cite{trotter,suzuki1,suzuki2}, the ITE is split into a product of local two-body Hamiltonians $\hat{H}_{ij}$ ($\hat{H}=\sum_{\langle i,j \rangle}\hat{H}_{ij}$) with nearest-neighbor interactions
\begin{eqnarray}
    e^{-T\hat{H}} \left| \Psi_{0} \right\rangle = \left[ \left( \prod_{\langle i,j \rangle} e^{-\tau\hat{H}_{ij}} \right)^{N_{\tau}} + O(\tau) \right] \left| \Psi_{0} \right\rangle, \label{tsdeco}
\end{eqnarray}
where $N_{\tau}=T/\tau$ is the number of ITE steps.
We neglect $O(\tau)$ by setting $\tau$ as a sufficiently small value, and iterate the ITE steps until it converges.
In practice, after multiplying $e^{-\tau\hat{H}_{ij}}$, we truncate the bond dimension by the simple update method \cite{vidal,jiang}, so that this remains $D$ throughout the ITE steps.
Note that we should start the ITE steps from a variety of possible initial conditions so as to reduce the influence of being trapped in a local minimum.

We then derive the expectation value of a physical quantity $\hat{O}$ defined as $\langle \hat{O} \rangle \equiv \langle \Psi | \hat{O} | \Psi \rangle/\langle \Psi | \Psi \rangle$.
By contracting indices $s_{i}$ in Eq. \eqref{peps2}, $\langle \Psi | \hat{O} | \Psi \rangle$ and $\langle \Psi | \Psi \rangle$ can also be expressed as an infinite tensor network with the bond dimension $D^{2}$.
We accurately approximated this as a finite one with the help of the corner transfer matrix renormalization group (CTMRG) method \cite{baxter1,baxter2,nishino2,orus2,phien,lee3} for an arbitrary unit cell structure \cite{corboz2}.
The accuracy of the CTMRG is determined by the bond dimension $\chi$ of the corner matrices and the edge tensors.
We choose $\chi$ as $\chi\propto D^{2}$. 
%
\begin{figure}[t]
    \begin{center}
    \includegraphics[keepaspectratio,scale=0.500]{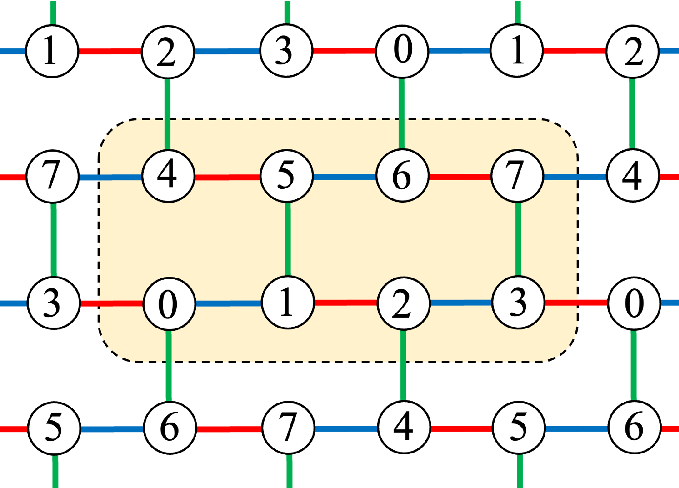}
    \end{center}
        \caption{Eight sublattices in a unit cell with periodic boundary conditions.
        The blue, red and green lines denote bonds of local tensors along $x$-, $y$-, and $z$-directions of the Kitaev term respectively.}
        \label{fig:sublattice}
\end{figure}

In the present iPEPS simulations, we adopt the tensor- network library TeNeS \cite{tenes1,tenes2}.
We define the iPEPS in a unit cell with eight sublattices (see Fig. \ref{fig:sublattice}).
Each circle denotes a local tensor $\psi^{s_{i}}_{l_{i}m_{i}n_{i}}$ in Eq. \eqref{peps2}.
As for the ITE, we set $\tau=10^{-2}$ and $N_{\tau}=10^{4}$ so that the ITE well converges.
Based on the results in Ref. \cite{pohle}, we prepare seven initial conditions for the ITE: AFM, FM, stripy, zigzag, and FQ states in addition to antiferro- and ferro-loop gas states (LGSs) \cite{lee,lee2} which qualitatively capture the nature of  AKSL and FKSL states respectively.
At a single parameter point $(\theta,\phi)$, we perform the ITE under these seven initial conditions.
We then employ a state where the energy becomes the lowest.
Considering the convergence of energy at Kitaev limits in Ref. \cite{lee2}, we set the bond dimensions as $(D,\chi)=(8,64)$ or $(8,128)$, described in the next section in more detail.

\section{Results}
\label{sec:results}
In this section, we give numerical results.
Figure \ref{fig:phase_sl} shows the close-up of the phase diagram Fig. \ref{fig:phase_etc}(a) in the vicinity of the Kitaev limits ($\theta=0$ and $\pi$).
We explain these phase diagrams in Sec. \ref{subsec:phase}.
In Secs. \ref{subsec:fksl} and \ref{subsec:aksl}, we describe how to detect different phases and in the vicinity of ferro- and antiferro-Kitaev limits respectively.

As for the bond dimensions of iPEPS, we adopt $(D,\chi)=(8,128)$ in the black squared region of Fig. \ref{fig:phase_sl}(a) and when determining the AKSL--FQ boundaries in Fig. \ref{fig:phase_sl}(b), so that the CTMRG well converges.
In the other parameter regions, we adopt $(D,\chi)=(8,64)$.

\subsection{Ground-state phase diagram}
\label{subsec:phase}
\begin{figure}[t]
    \begin{center}
    \includegraphics[keepaspectratio,scale=0.50]{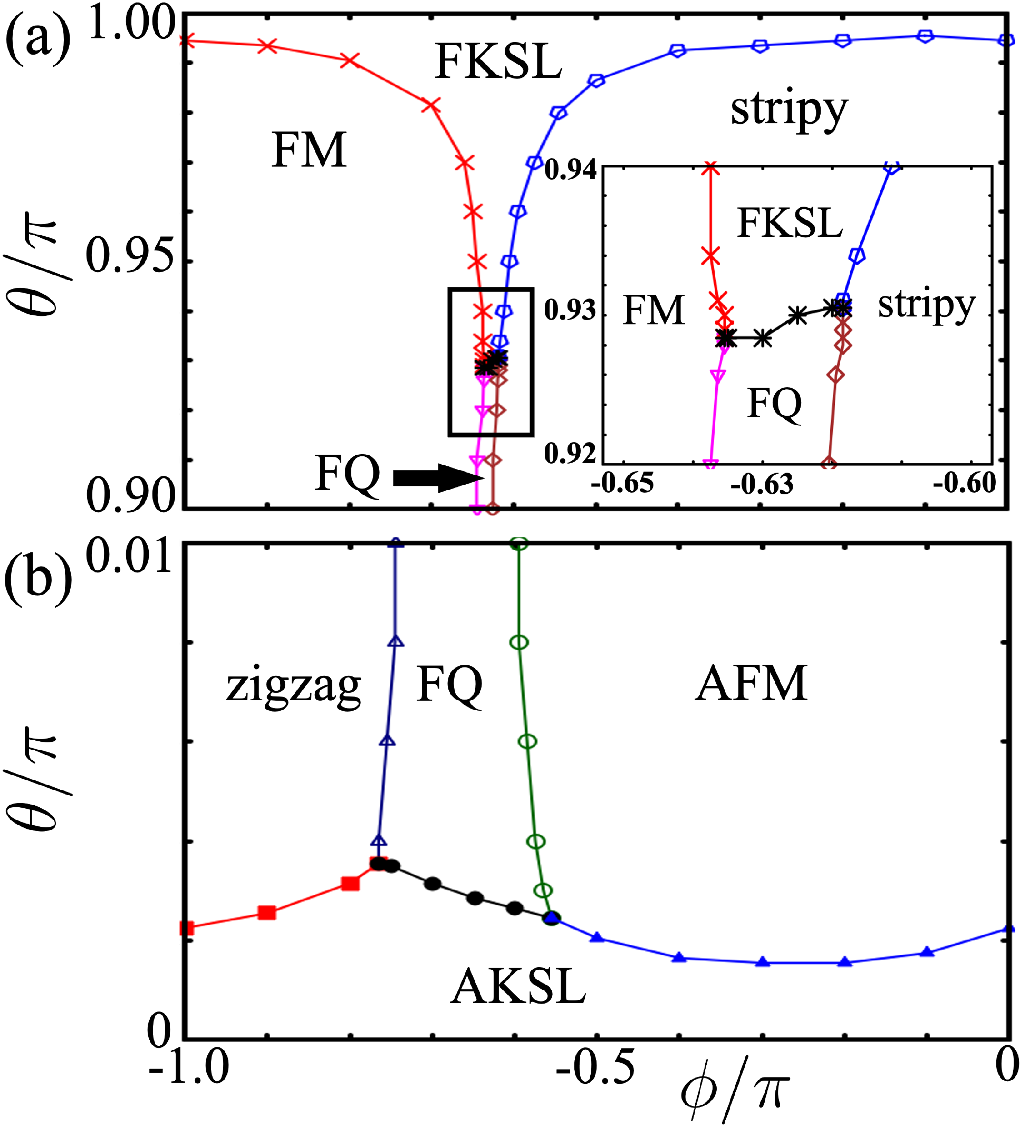}
    \end{center}
        \caption{Phase diagram of the BBQ-K model: (a) the vicinity of the ferro-Kitaev limit ($\theta=\pi$) with the inset closing up the black squared region, and (b) the vicinity of the antiferro-Kitaev limit ($\theta=0$).}
        \label{fig:phase_sl}
\end{figure}

In Fig. \ref{fig:phase_sl}, there are phase transitions between the KSL and the four spin-dipolar ordered phases.
Considering the Kitaev--Heisenberg model ($\phi/\pi=-1.0$ or $0.0$), our results of boundaries between these phases are consistent with the density matrix renormalization group (DMRG) calculation \cite{dong}.
More interestingly, we newly discover the direct KSL--FQ transitions appear in the vicinity of both Kitaev limits.
This result is significant since these transitions are in principle impossible to detect until one accurately capture quantum fluctuations and entanglements by iPEPS.

In the vicinity of the ferro-Kitaev limit ($\theta=\pi$), we confirm that the FKSL phase drastically gets extended around the region of $-0.65\le\phi/\pi\le-0.6$ [see Fig. \ref{fig:phase_sl}(a)].
In this region, the ferro-quadrupolar term of the BBQ model is dominant compared to the Heisenberg one.
Especially at $\phi=\arctan{(2)}$ ($\approx-0.648\pi$), the BBQ-K model Eq. \eqref{bbqkij} can be rewritten without the the Heisenberg term as
\begin{eqnarray}
    \hat{H}^{\langle i,j \rangle_{\gamma}}_{\mathrm{BBQ-K}} &=& -\frac{\sin \theta}{\sqrt{5}} \left( \hat{\bm{Q}}_{i} \cdot \hat{\bm{Q}}_{j} + \frac{8}{3} \right) +\cos\theta \hat{S}^{\gamma}_{i} \hat{S}^{\gamma}_{j}. \label{bbqkq}
\end{eqnarray}
We thus interpret it as meaning that the FKSL phase is robust against perturbations from ferro-quadrupolar interactions.
Moreover, there is not the direct FM--stripy transition unlike Ref. \cite{pohle} since KSL or FQ phase exists between FM and stripy phases.
We find that this intermediate FQ phase appears since it is more robust against the Heisenberg interaction than one in Ref. \cite{pohle}, as iPEPS accurately capture quantum entanglements.

In the vicinity of the antiferro-Kitaev limit ($\theta=0$) shown in Fig. \ref{fig:phase_sl}(b), we also discover that the FQ phase is also more robust against the antiferro-Kitaev interaction than one in Ref. \cite{pohle}, while the AKSL phase is vulnerable to both Heisenberg and ferro-quadrupolar interactions.
Note that the FQ phase get extended when the antiferro-Kitaev interaction becomes dominant, which shrinks in Ref. \cite{pohle}.
Also, there is not the direct zigzag--AFM transition since the FQ phase get between these two phases.
We reveal novel quantum properties caused by the competition between antiferro-Kitaev and ferro-quadrupolar interactions by iPEPS.

\subsection{Vicinity of the ferro-Kitaev limit}
\label{subsec:fksl}
In this subsection, we explain how to calculate order parameters for determining phases and boundaries with the emphasis on the vicinity of ferro-Kitaev limit ($\theta=\pi$).
In later subsubsections, we show our numerical results in the cases of the FQ--FKSL and FM--FQ--stripy transitions.

\subsubsection{FQ--FKSL transition}
In this subsubsection, we show our numerical results illustrating the FQ--FKSL transition.
We derive some order parameters to determine phases and the boundary when tuning the parameter $\theta$ and fixing $\phi$ at $\phi=-0.63\pi$.

\begin{figure}[t]
    \begin{center}
    \includegraphics[keepaspectratio,scale=1.08]{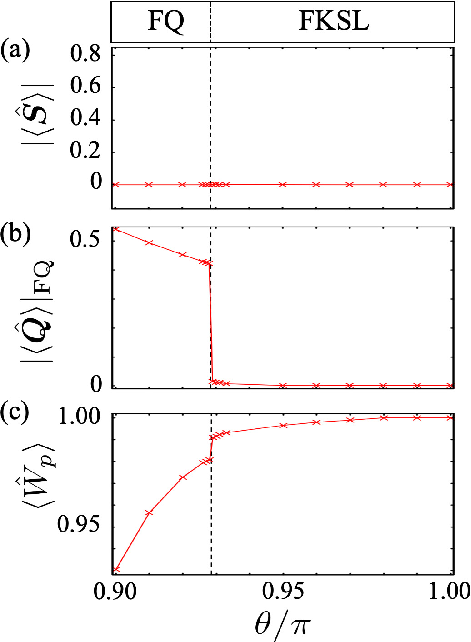}
    \end{center}
        \caption{Numerical results as a function of $\theta$ when fixing $\phi$ at $\phi=-0.63\pi$: (a) norm of spin $| \langle \hat{\bm{S}} \rangle |$, (b) norm of ferro-quadrupolar order parameter $| \langle \hat{\bm{Q}} \rangle |_{\mathrm{FQ}}$, (c) flux $\langle \hat{W}_{p} \rangle$.
        The dashed line is the phase boundary between FQ phase and FKSL phase: $\theta/\pi\approx 0.9285$.
        Here, we set $(D,\chi)=(8,128)$.}
        \label{fig:p-0.63_fsl}
\end{figure}

Here, we calculate the norm of the spin dipolar moment to investigate magnetic orders [see Fig. \ref{fig:p-0.63_fsl}(a)], defined as
\begin{eqnarray}
    | \langle \hat{\bm{S}} \rangle | \equiv \frac{1}{8} \sum_{i \in \{0,\cdots,7\}} \sqrt{\langle \hat{S}^{x}_{i} \rangle^{2}+\langle \hat{S}^{y}_{i} \rangle^{2}+\langle \hat{S}^{z}_{i} \rangle^{2}}. \label{abs}
\end{eqnarray}
In Eq. \eqref{abs}, $i \in \{0,\cdots,7\}$ means the eight sublattices in the unit cell shown in Fig. \ref{fig:sublattice}.
$| \langle \hat{\bm{S}} \rangle |$ takes a finite value if the system has any magnetic orders, and 0 otherwise.
As shown in Fig. \ref{fig:p-0.63_fsl}(a), $| \langle \hat{\bm{S}} \rangle |$ stays 0 in the whole range of $0.90\le\theta/\pi\le1.0$, which corresponds to characteristics of FQ and FKSL phases without spin-dipolar order.

Next, we investigate the ferro-order of the quadrupolar moment $| \langle \hat{\bm{Q}} \rangle |_{\mathrm{FQ}}$, defined as
\begin{eqnarray}
    | \langle \hat{\bm{Q}} \rangle |_{\mathrm{FQ}} \equiv \sqrt{\sum_{\gamma=1}^{5} \left( \frac{1}{8} \sum_{i \in \{0,\cdots,7\}} \langle \hat{Q}_{i}^{\gamma} \rangle \right)^{2}}. \label{absfq}
\end{eqnarray}
$| \langle \hat{\bm{Q}} \rangle |_{\mathrm{FQ}}$ takes finite values in the case of ferro-quadrupolar ordered state, and 0 in the other cases.
We confirm a sharp jump of $| \langle \hat{\bm{Q}} \rangle |_{\mathrm{FQ}}$ [see Fig. \ref{fig:p-0.63_fsl}(b)] at $\theta/\pi\approx 0.9285$, and the left-side area $0.9\le\theta/\pi\lesssim 0.9285$ is the FQ phase.
$| \langle \hat{\bm{Q}} \rangle |_{\mathrm{FQ}}$ becomes small but finite in the right side area near the boundary $\theta/\pi\approx 0.9285$.
However, we expect that this approaches to 0 if we increase the bond dimension $D$.

We examine the vortex freeness in this parameter region by calculating the expectation value of the flux $\hat{W}_{p}$ in Eq. \eqref{flux}, defined as $\langle \hat{W}_{p} \rangle \equiv \frac{1}{4} \sum_{j=1}^{4} \langle \hat{W}_{p_{j}} \rangle$.
$\hat{W}_{p_{j}}$ means the flux on the plaquette $p_{j}$ in the Fig. \ref{fig:sublattice}, defined as 
\begin{eqnarray*}
    &&\hat{W}_{p_{1}} \equiv \hat{U}_{0}^{z}\hat{U}_{1}^{y}\hat{U}_{5}^{x}\hat{U}_{4}^{z}\hat{U}_{7}^{y}\hat{U}_{3}^{x}, \hspace{0.5em} 
    \hat{W}_{p_{2}} \equiv \hat{U}_{2}^{z}\hat{U}_{3}^{y}\hat{U}_{7}^{x}\hat{U}_{6}^{z}\hat{U}_{5}^{y}\hat{U}_{1}^{x},  \\
    &&\hat{W}_{p_{3}} \equiv \hat{U}_{5}^{z}\hat{U}_{6}^{y}\hat{U}_{0}^{x}\hat{U}_{3}^{z}\hat{U}_{2}^{y}\hat{U}_{4}^{x}, \hspace{0.5em} 
    \hat{W}_{p_{4}} \equiv \hat{U}_{7}^{z}\hat{U}_{4}^{y}\hat{U}_{2}^{x}\hat{U}_{1}^{z}\hat{U}_{0}^{y}\hat{U}_{6}^{x}. 
\end{eqnarray*}
Figure. \ref{fig:p-0.63_fsl}(c) shows a small jump of $\langle \hat{W}_{p} \rangle$ at the boundary $\theta/\pi=0.9285$.
In the right-side area $0.9285\lesssim\theta/\pi\le1.0$, the FKSL phase is extended without any other intermediate phases since $| \langle \hat{\bm{S}} \rangle |=| \langle \hat{\bm{Q}} \rangle |_{\mathrm{FQ}}\approx0$ and $\langle \hat{W}_{p} \rangle\approx 1$.
Interestingly, $\langle \hat{W}_{p} \rangle$ takes the value near to 1 even in the FQ phase, which means that the FQ phase shows the vortex freeness strongly influenced by the FKSL phase.


\subsubsection{FM--FQ--stripy transition}
\begin{figure}[t]
    \begin{center}
    \includegraphics[keepaspectratio,scale=1.12]{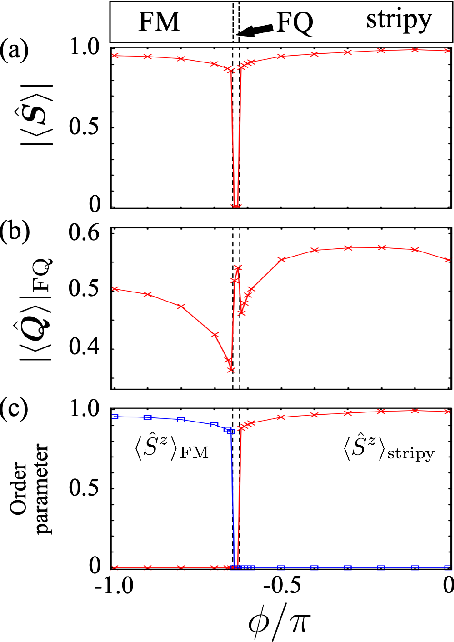}
    \end{center}
        \caption{Numerical results as a function of $\phi$ when fixing $\theta$ at $\theta=0.9\pi$: (a) norm of spin $| \langle \hat{\bm{S}} \rangle |$, (b) norm of ferro-quadrupolar order parameter $| \langle \hat{\bm{Q}} \rangle |_{\mathrm{FQ}}$, (c) order parameters of FM and stripy states: $\langle \hat{S}^{z} \rangle_{\mathrm{FM}}$ and $\langle \hat{S}^{z} \rangle_{\mathrm{stripy}}$ respectively.
        Two dashed lines are the phase boundaries of the three phases: $\phi/\pi\approx -0.645$ and $-0.625$.
        Here, we set $(D,\chi)=(8,64)$.}
        \label{fig:t0.9}
\end{figure}
In this subsubsection, we illustrate results of the FM--FQ--stripy transition by calculating some order parameters when tuning the parameter $\phi$ and fixing $\theta$ at $\theta=0.9\pi$.
We calculate the norms of the spin dipolar and quadrupolar moments, $| \langle \hat{\bm{S}} \rangle |$ and $| \langle \hat{\bm{Q}} \rangle |_{\mathrm{FQ}}$ respectively, to see magnetic orders [see Figs. \ref{fig:t0.9}(a) and \ref{fig:t0.9}(b)].
As shown in Fig. \ref{fig:t0.9}(a), $| \langle \hat{\bm{S}} \rangle |$ remains finite except for a narrow middle region where it suddenly drops to 0.
Also, we confirm an increase of $| \langle \hat{\bm{Q}} \rangle |_{\mathrm{FQ}}$ in this narrow region [see Fig. \ref{fig:t0.9}(b)].
According to these results, we determine that this middle region is the FQ phase whose boundaries are $\phi/\pi \approx -0.645$ and $-0.625$.

Next, we define two order parameters to determine magnetic orders.
We define an FM order parameter as
\begin{eqnarray}
    \langle \hat{S}^{z} \rangle_{\mathrm{FM}} &\equiv& \frac{1}{8} \left| \sum_{i \in \{0,\cdots,7\}} \langle \hat{S}^{z}_{i} \rangle \right|. \label{ofm}
\end{eqnarray}
We write a stripy order parameter as $\langle \hat{S}^{z} \rangle_{\mathrm{stripy}} \equiv \max\left\{ \langle \hat{S}^{z} \rangle_{\mathrm{stripy1}}, \langle \hat{S}^{z} \rangle_{\mathrm{stripy2}}, \langle \hat{S}^{z} \rangle_{\mathrm{stripy3}} \right\}$ and 
\begin{eqnarray}
    \langle \hat{S}^{z} \rangle_{\mathrm{stripy}a} \equiv \frac{1}{8} \left| \left(  \sum_{i \in A_{a}} -  \sum_{i \in B_{a}} \right) \langle \hat{S}^{z}_{i} \rangle \right|, \label{ostripy}
\end{eqnarray}
where $A_{a}$ and $B_{a}$ ($a=1,2,3$) are sets of sites in the unit cell Fig. \ref{fig:sublattice}, defined as $A_{1}=\{0,2,4,6\}$, $B_{1}=\{1,3,5,7\}$, $A_{2}=\{0,1,4,7\}$, $B_{2}=\{2,3,5,6\}$, $A_{3}=\{0,3,4,5\}$, $B_{3}=\{1,2,6,7\}$.
Namely, $\langle \hat{S}^{z} \rangle_{\mathrm{stripy}}$ becomes a finite value if spins take any stripy ordered configurations shown in Fig. \ref{fig:phase_etc}(b), and 0 otherwise.
From Fig. \ref{fig:t0.9}(c), we find that the left-side magnetic region is the FM phase, while the right-side one is the stripy phase.
Note that the direct FM--stripy does not appear unlike Ref. \cite{pohle} since the FQ phase get between them, which can only be detected by appropriately illustrating quantum entanglements ascribed for the ferro-Kitaev interaction utilizing the iPEPS.

\subsection{Vicinity of the antiferro-Kitaev limit}
\label{subsec:aksl}
In this subsection, we explain how to detect phases by calculating order parameters with the emphasis on the vicinity of antiferro-Kitaev limit ($\theta=0$).
In later subsubsections, we show our numerical results in the cases of the AKSL--FQ and zigzag--FQ--AFM transitions.

\subsubsection{AKSL--FQ transition}
In this subsection, we show our numerical results illustrating the AKSL--FQ transition, in the vicinity of the antiferro-Kitaev limit ($\theta=0$).
We show the physical quantities by varying the parameter $\theta$, while keeping $\phi$ fixed at $\arctan(2)$.
\begin{figure}[t]
    \begin{center}
    \includegraphics[keepaspectratio,scale=1.09]{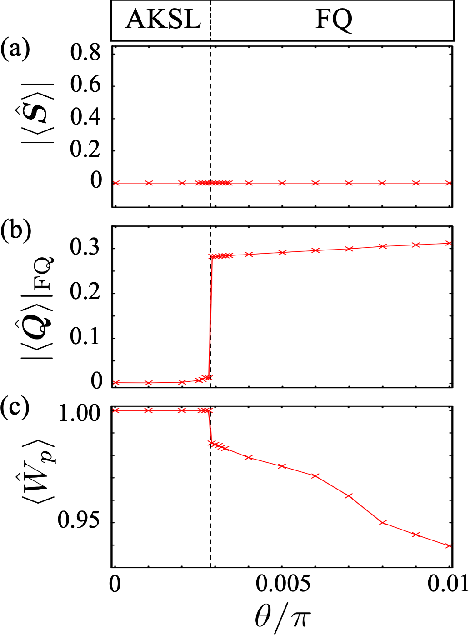}
    \end{center}
        \caption{Numerical results as a function of $\theta$ when fixing $\phi$ at $\phi=\arctan{(2)}\approx-0.648\pi$: (a) norm of spin $| \langle \hat{\bm{S}} \rangle |$, (b) norm of ferro-quadrupolar order parameter $| \langle \hat{\bm{Q}} \rangle |_{\mathrm{FQ}}$, (c) flux $\langle \hat{W}_{p} \rangle$.
        The dashed line is the phase boundary between AKSL phase and FQ phase: $\theta/\pi\approx 0.00285$.
        Here, we set $(D,\chi)=(8,128)$.}
        \label{fig:p-0.648_afsl}
\end{figure}

To begin with, we investigate dipolar and quadrupolar orders as shown in Figs. \ref{fig:p-0.648_afsl}(a) and \ref{fig:p-0.648_afsl}(b).
Although $| \langle \hat{\bm{S}} \rangle |$ stays 0 in the whole range of $0\le\theta/\pi\le0.01$, we confirm that $| \langle \hat{\bm{Q}} \rangle |_{\mathrm{FQ}}$ shows a sharp jump at $\theta/\pi\approx0.00285$.
$| \langle \hat{\bm{Q}} \rangle |_{\mathrm{FQ}}$ becomes small but finite in the right side area near the boundary $\theta/\pi\approx 0.00285$.
However, we expect that this approaches to 0 if we increase the bond dimension $D$.
We therefore determine the right-side region with the finite $| \langle \hat{\bm{Q}} \rangle |_{\mathrm{FQ}}$ is the FQ phase.
We then examine the flux $\langle \hat{W}_{p} \rangle$ to see the vortex freeness.
Figure \ref{fig:p-0.648_afsl}(c) also shows a small jump of $\langle \hat{W}_{p} \rangle$ at the boundary $\theta/\pi=0.00285$.
From this result, we determine the left-side area ($0.0\lesssim\theta/\pi\le0.00285$) is the AKSL phase since $| \langle \hat{\bm{S}} \rangle |=| \langle \hat{\bm{Q}} \rangle |_{\mathrm{FQ}}\approx0$ and $\langle \hat{W}_{p} \rangle\approx 1$.
Interestingly, similar to the case near the antiferro-Kitaev limit, $\langle \hat{W}_{p} \rangle$ takes the value near to 1 even in the FQ phase, showing the vortex freeness of the FQ phase strongly influenced by the AKSL phase.

\subsubsection{zigzag--FQ--AFM transition}
\begin{figure}[t]
    \begin{center}
    \includegraphics[keepaspectratio,scale=1.12]{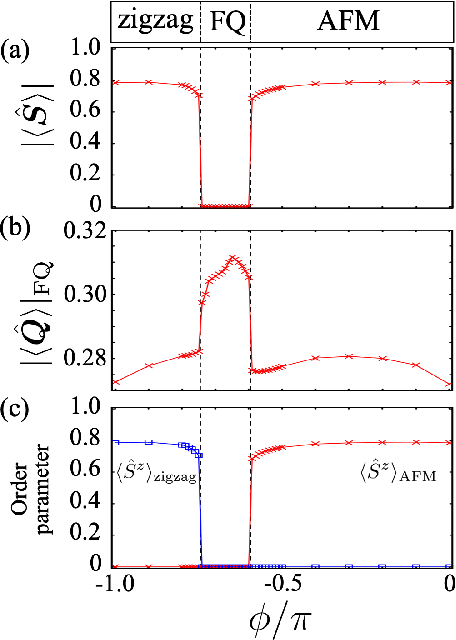}
    \end{center}
        \caption{Numerical results as a function of $\phi$ when fixing $\theta$ at $\theta=0.01\pi$: (a) norm of spin $| \langle \hat{\bm{S}} \rangle |$, (b) norm of ferro-quadrupolar order parameter $| \langle \hat{\bm{Q}} \rangle |_{\mathrm{FQ}}$, (c) order parameters of zigzag and AFM states: $\langle \hat{S}^{z} \rangle_{\mathrm{zigzag}}$ and $\langle \hat{S}^{z} \rangle_{\mathrm{AFM}}$ respectively.
        Two dashed lines are the phase boundaries of the three phases: $\phi/\pi\approx -0.745$ and $-0.595$.
        Here, we set $(D,\chi)=(8,64)$.}
        \label{fig:t0.01}
\end{figure}

In this subsection, we explain results of the zigzag--FQ--AFM transition by calculating some order parameters when tuning the parameter $\phi$ and fixing $\theta$ at $\theta=0.01\pi$.
We calculate $| \langle \hat{\bm{S}} \rangle |$ and $| \langle \hat{\bm{Q}} \rangle |_{\mathrm{FQ}}$ to see magnetic orders [see Figs. \ref{fig:t0.01}(a) and \ref{fig:t0.01}(b)].
As shown in Fig. \ref{fig:t0.01}(a), $| \langle \hat{\bm{S}} \rangle |$ remains finite values except for an intermediate extended region $-0.745 \le \phi/\pi \le -0.595$, where it suddenly drops to 0.
We also confirm a sharp jump of $| \langle \hat{\bm{Q}} \rangle |_{\mathrm{FQ}}$ in this region [see Fig. \ref{fig:t0.01}(b)].
We thus determine that this intermediate region is the FQ phase.

Next, we derive two order parameters to determine what phases the magnetic ordered regions are.
We defined an AFM order parameter as
\begin{eqnarray}
    \langle \hat{S}^{z} \rangle_{\mathrm{AFM}} &\equiv& \frac{1}{8} \left| \left(  \sum_{i \in \{0,2,5,7\}} -  \sum_{i \in \{1,3,4,6\}} \right)\langle \hat{S}^{z}_{i} \rangle \right|. \label{oafm}
\end{eqnarray}
We defined a zigzag order parameter as $\langle \hat{S}^{z} \rangle_{\mathrm{zigzag}} \equiv \max\left\{ \langle \hat{S}^{z} \rangle_{\mathrm{zigzag1}}, \langle \hat{S}^{z} \rangle_{\mathrm{zigzag2}}, \langle \hat{S}^{z} \rangle_{\mathrm{zigzag3}} \right\}$ and
\begin{eqnarray}
    \langle \hat{S}^{z} \rangle_{\mathrm{zigzag}a} \equiv \frac{1}{8} \left| \left(  \sum_{i \in C_{a}} -  \sum_{i \in D_{a}} \right) \langle \hat{S}^{z}_{i} \rangle \right|,\label{ozigzag}
\end{eqnarray}
where $C_{a}$ and $D_{a}$ ($a=1,2,3$) are sets of sites in the unit cell Fig. \ref{fig:sublattice}, defined as $C_{1}=\{0,1,2,3\}$, $D_{1}=\{4,5,6,7\}$, $C_{2}=\{0,3,6,7\}$, $D_{2}=\{1,2,4,5\}$, $C_{3}=\{0,1,5,6\}$, $D_{3}=\{2,3,4,7\}$.
Namely, $\langle \hat{S}^{z} \rangle_{\mathrm{zigzag}}$ becomes a finite value if spins take any zigzag ordered configurations shown in Fig. \ref{fig:phase_etc}(b), and 0 otherwise.
From Fig. \ref{fig:t0.01}(c), we find that the left-side magnetic region is the zigzag phase, while the right-side one is the AFM phase.
Unlike Ref. \cite{pohle}, according to Figs. \ref{fig:phase_sl}(c) and \ref{fig:t0.01}, the direct zigzag--AFM does not occur since the FQ phase enters between them, which can only be discovered by appropriately illustrating quantum entanglement by iPEPS.

\section{Conclusion and discussion}
\label{sec:con}

In order to probe quantum phases and phase transitions ascribed for the competition between the KSL and the SN, we numerically investigate the BBQ-K model Eq. \eqref{bbqk}, which was originally proposed in Ref. \cite{pohle}.
Utilizing the 2D iPEPS, we take into account quantum entanglements among spins ignored in Ref. \cite{pohle}.
Then, we could represent quantum properties of the KSL more precisely.
As a result, we succeed in constructing the phase diagram (see Figs. \ref{fig:phase_etc}(a) and \ref{fig:phase_sl}), including extended KSL phases which could not, in principle, be captured by the semi-classical variational calculations in Ref. \cite{pohle}.
More specifically, we result in discovering several new properties:
(1) We detect the direct KSL-FQ phase transitions (see Fig. \ref{fig:phase_sl}).
(2) The FKSL phase is stabilized under the almost pure ferro-quadrupolar perturbation [see Fig. \ref{fig:phase_sl}(a)].
(3) The FQ phase is more robust against the Heisenberg interaction than one obtained in Ref. \cite{pohle}, and it gets extended when the antiferro-Kitaev interaction becomes dominant [see Fig. \ref{fig:phase_sl}(b)]. 
Note that it is, as far as we know, the first time that a direct phase transition between a quantum spin liquid phase and an SN phase is discovered in higher spin-$S$ systems, although similar results were obtained \cite{nasu} in the spin-1/2 system as for a bond-nematic state.
\begin{figure}[t]
    \begin{center}
    \includegraphics[keepaspectratio,scale=1.06]{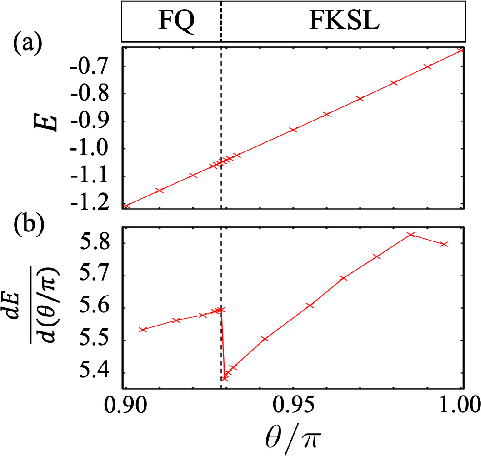}
    \end{center}
        \caption{Numerical results as a function of $\theta$ when fixing $\phi$ at $\phi=-0.63\pi$: (a) energy $E$, and (b) first derivative of the energy.
        Here, we set $(D,\chi)=(8,128)$.}
        \label{fig:p-0.63_fsl_ene}
\end{figure}

As an open problem, we could not determine whether the phase transitions seen in this paper are the second-ordered or first-ordered transition.
As shown in Fig. \ref{fig:p-0.63_fsl_ene}, we calculate the energy and its first derivative to determine the order of the phase transitions.
As for the first derivative, we consider the numerical differentiation as 
\begin{eqnarray}
    \left. \frac{dE}{d\theta} \right|_{\theta=\theta^{\prime}-\frac{\Delta \theta}{2}} \approx \frac{E(\theta^{\prime}) - E(\theta^{\prime}-\Delta\theta)}{\Delta \theta},
\end{eqnarray}
based on the numerical data at $\theta^{\prime}$ and $\theta^{\prime} - \Delta \theta$.
We here confirm that the first derivative shows a jump describing the first-ordered transition in the case of $\phi=-0.63\pi$ near the ferro-Kitaev limit [see Fig. \ref{fig:p-0.63_fsl_ene}(b)].
Similar results were obtained in the other cases listed in Sec. \ref{sec:results}.
However, these results seem to be not reliable. 
The ITE utilizing the simple update could not appropriately capture the long-range correlation, and thus, the ground states obtained by iPEPS strongly depend on initial states.
Therefore, we could not rule out the possibility of the second-ordered transition.

Another open problem is the nature of the phase diagram for the positive quadrupolar interaction $J_{2}>0$ ($0\le\phi\le\pi$), where Ref. \cite{pohle} proposed several exotic phases.
In this region, the AFQ order appeared in classical spin-1 systems \cite{chen,papanicolaou} is melted by strong quantum fluctuation emerged from the competition between the Heisenberg and the positive biquadratic (quadrupolar) terms.
Instead, there appears \cite{zhao,corboz} plaquette valence-bond solid state characterized by the plaquette order breaking translational symmetry.
Investigating the interplay between the KSL and the positive quadrupolar interaction seems to be an interesting near future task, which may require more computational cost than the case of this paper, namely $D>8$, according to the numerical work with the SU(3) Heisenberg model \cite{corboz}.
Since the positive quadrupolar interaction is expected in materials with orbital degeneracy \cite{yoshimori,hoffmann,soni}, investigating the models with $J_{2} > 0$ might be relevant for understanding the nature of these materials.

One of future tasks is an investigation of the low-energy excitation, which offer clear guidelines for experiments with inelastic neutron scattering.
This excitation is expected to be caused by waves of local order parameters in the BBQ-K model.
For describing this excitation, one should utilize the SU(3) flavor-wave theory \cite{papanicolaou2,joshi,lauchli1}, since the conventional SU(2) spin-wave theory fails to illustrate quadrupolar fluctuations ascribed for the biquadratic term \cite{wysocki,stanek,hu,yu,bilbao}.

Finally, we discuss implications of our results for experiments.
According to our numerical results in Fig. \ref{fig:phase_sl}(a), the FKSL phase can realize even with a relatively weak ferro-Kitaev interaction, if the pure ferro-quadrupolar interaction exists.
Therefore, our phase diagrams and results in Sec. \ref{subsec:fksl} offer clues for experiments searching for materials showing FKSL state.
On the other hand, the AKSL phase is vulnerable to the ferro-quadrupolar interaction [see Fig. \ref{fig:phase_sl}(b)], unlike the FKSL phase.
This result may offer clues for experiments with candidate antiferro-Kitaev materials like $A_{3}$Ni$_{2}X$O$_{6}$ ($A=$ Li, Na, $X=$ Bi, Sb) \cite{stavropoulos} to detect the AKSL--FQ transition.
We desire that our results will stimulate further experimental and theoretical studies on the interplay between the KSL and the SN states.

\section{Acknowledgments}
We appreciate R. Pohle and Y. Motome for helpful discussions.
Our work is financially supported by Japan Society for the Promotion of Science KAKENHI, Grant Nos. 20H00122, 22H01179, 22K18682, 23H03818, and by the Center of Innovations for Sustainable Quantum AI (JST Grant Number JPMJPF2221).
T. O. acknowledges the support from the Endowed Project for Quantum Software Research and Education, The University of Tokyo (https://qsw.phys.s.u-tokyo.ac.jp/).
We utilize the tensor-network software, named TeNeS \cite{tenes1,tenes2} for iPEPS calculations.
Our numerical calculations was implemented by the Supercomputer Center of the Institute for Solid State Physics, the University of Tokyo.


%
\end{document}